\documentclass{iopart}
\usepackage{epsfig}
\usepackage{amssymb}

\usepackage{color}

\begin{document} 

\newcommand{\vk}{{\vec k}} 
\newcommand{\vK}{{\vec K}}  
\newcommand{\vb}{{\vec b}}  
\newcommand{\vp}{{\vec p}}  
\newcommand{\vq}{{\vec q}}  
\newcommand{\vQ}{{\vec Q}} 
\newcommand{\vx}{{\vec x}} 
\newcommand{\vv}{{\vec v}} 
\newcommand{\vh}{{\hat{v}}} 
\newcommand{\cO}{{\cal O}}
\newcommand{\be}{\begin{equation}} 
\newcommand{\ee}{\end{equation}}  
\newcommand{\half}{{\textstyle\frac{1}{2}}}  
\newcommand{\gton}{\mathrel{\lower.9ex \hbox{$\stackrel{\displaystyle 
>}{\sim}$}}}  
\newcommand{\lton}{\mathrel{\lower.9ex \hbox{$\stackrel{\displaystyle 
<}{\sim}$}}}  
\newcommand{\ben}{\begin{enumerate}}  
\newcommand{\een}{\end{enumerate}} 
\newcommand{\bit}{\begin{itemize}}  
\newcommand{\eit}{\end{itemize}} 
\newcommand{\bc}{\begin{center}}  
\newcommand{\ec}{\end{center}} 
\newcommand{\bea}{\begin{eqnarray}}  
\newcommand{\eea}{\end{eqnarray}}

\title{Particle correlations at RHIC from parton coalescence dynamics -- first results}
 
\date{\today}
 
\author{D\'enes Moln\'ar}
\address{Department of Physics, The Ohio State University, Columbus, OH 43210}

\begin{abstract}
A new dynamical approach
that combines covariant parton transport theory
with hadronization channels via
parton coalescence and fragmentation
is applied to Au+Au at RHIC.
Basic consequences of the simple coalescence formulas, 
such as elliptic flow scaling and enhanced $p$/$\pi$ ratio,
turn out to be rather 
sensitive to the spacetime aspects of coalescence dynamics.
\end{abstract}



\section{Introduction}
Two exciting recent discoveries in $Au+Au$ reactions at $\sqrt{s_{NN}}=200$ 
GeV at RHIC were the lack of baryon suppression\cite{QM2004exp,STARboth,PHENIXnoBsupp,STARnoBsupp} 
in the region $2 < p_\perp < 5$ GeV
and the quark number scaling 
of elliptic flow\cite{QM2004exp,PHENIXv2scaling,STARboth}.
Parton coalescence\cite{QM2004th,HwaYang,Voloshincoal,texbudMtoB,dukeCoal,coalv2,charmv2} is currently the most promising
proposal to explain both phenomena.

In the coalescence model,
mesons(baryons) form via a fusion of two(three) 
quarks/antiquarks.
The simplest forms of the model
based on (\ref{coaleq}), or minor variations
of it,
were fairly successful in reproducing particle spectra 
at RHIC\cite{texbudMtoB,dukeCoal}
in terms of a common set of 
``reasonable'' constituent phasespace distributions 
over a ``reasonable'' hadronization hypersurface.
These parameters were consistent with the assumption of 
a deconfined quark-gluon plasma.
The scaling of elliptic flow $v_2(p_\perp)$ with constituent
number is also understood based on
these simple formulas\cite{coalv2,charmv2}.

Nevertheless, these earlier studies left several important questions open.
For example, it is known\cite{coalv2,charmv2} that (\ref{coaleq}) 
violates unitarity.
The yield in a given coalescence channel scales quadratically/cubically
with constituent number,
moreover, the same constituent contributes to several channels.
It is also unclear whether the extracted hadronization parameters
are consistent with any dynamical scenario.
In addition, the simplified form of constituent 
phasespace distributions assumed ignores several types of phasespace 
correlations that would be generated in a dynamical approach.

The goal of this study is to improve upon the above deficiencies
and study how the dynamics of parton coalescence affects correlation
observables at RHIC, such as elliptic flow.

\section{Dynamical coalescence approach}
\label{formalism}

Most of the parton coalescence formalism 
is based on studies of deuteron 
formation\cite{GFR,Dover,Nagle,Scheibl}.
In the framework of nonrelativistic (equal-time) $N$-body
quantum mechanics, the number of deuterons (of given momentum)
at time $t$ is 
$N_d(t) = N_{pairs} \Tr[\hat\rho_d\, \hat \rho^{(2)}(t)]$,
where $\hat \rho_d\equiv|\Phi_d\rangle\langle \Phi_d|$
is the deuteron density matrix,
$\hat \rho^{(2)}(t)$ is the projection of the density matrix 
$|\Psi^{(N)}(t)\rangle \langle \Psi^{(N)}(t)|$ of the system 
onto the deuteron (two-particle) subspace,
there are $N_{pairs}$ possible $n{-}p$ pairs,
while $\Phi_d$ and $\Psi^{(N)}(t)$ are the wave function of
the deuteron and the system.
One is interested in
the observed deuteron number $N_d(t{\to}\infty)$
but unfortunately cannot solve the full $N$-body problem.

One approximation is to postulate that interactions {\em cease suddenly}
at some time $t_{f}$, 
and approximate $\hat \rho^{(2)}$ in the Wigner representation 
as the product of classical phase space densities $f_n \times f_p$ 
on the $t=t_{f}$ hypersurface.
The approximation, at best, is valid for weak bound states
and ignores genuine two-particle correlations.
Applied to meson formation $q\bar q\to M$, one obtains
the ``simple coalescence formula''
\be
\fl
\frac{dN_M(\vp)}{d^3p} \! =\! g_M \! 
\int\! d^3 x_1 d^3 p_1 d^3 x_2 d^3 p_2\, W_M(\Delta \vx,\Delta \vp)\,
f_q(\vp_1,\vx_1) f_{\bar q}(\vp_2,\vx_2)\, 
\delta^3(\vp{-}\vp_1{-}\vp_2)
\label{coaleq}
\ee
with $\Delta \vx \equiv \vx_1 - \vx_2$, $\Delta \vp \equiv \vp_1 - \vp_2$,
and the meson Wigner function
$W_M(\vx,\vp) \equiv \int d^3 b\, \exp[-i\vb\, \!\vp\,]
\Phi_M^*(\vx-\vb/2) \Phi_M(\vx+\vb/2)$. The degeneracy factor $g_M$ 
takes care of quantum numbers (flavor, spin, color).
The formula for baryons involves a triple phasespace integral and
the baryon Wigner function.
The generalization to arbitrary 3D hadronization hypersurfaces is 
straightforward\cite{Dover,Scheibl}.

One difficulty with (\ref{coaleq}) is
the proper choice of the hypersurface. 
While quantum mechanics gives a constant deuteron number 
at {\em any time} after freezeout, 
(\ref{coaleq}) decreases\cite{GFR,Nagle} with $t_f$
for a free streaming
distribution.
Also, transport approaches (i.e., self-consistent freezeout) 
yield diffuse freezeout 
distributions\cite{adrianFO,diffuseFO,ziweiFO} in full 4D spacetime.
These problems have been addressed in \cite{GFR}
by Gyulassy, Frankel and Remler (GFR), where they derived a way to interface
transport models and the coalescence formalism (in the weak binding limit).

The GFR result is essentially the same as (\ref{coaleq}),
except that the weight $W_M$ is evaluated using the {\em freezeout spacetime
points} $(t_1,\vx_1)$, $(t_2,\vx_2)$ of each constituent pair.
When taking $\Delta \vx$, the earlier particle needs to be propagated to the
time of the {\em later} one, resulting in an extra term, e.g.,
$\Delta \vx = \vx_1 - \vx_2 + (t_2 - t_1) \vv_1$ if $t_1 < t_2$.
The origin of this correction is that a weak bound state can 
only survive if none of its constituents have any further interactions.
The generalization to baryons involves propagation to the {\em latest} 
of the three freezeout times.

To investigate coalescence dynamics at RHIC,
we implanted GFR into 
covariant parton transport theory\cite{ZPC,MPC,v2,diffuseFO}.
First the parton ($g,u,d,s,\bar u, \bar d, \bar s$) evolution
was computed until freezeout using Molnar's Parton Cascade (MPC) 1.6.7 \cite{MPC}.
For simplicity, only $2\to 2$ processes were considered,
with Debye-screened parton cross sections
$d\sigma/dt \propto 1/(t-\mu_D^2)^2$, $\mu_D\approx 0.7$ GeV.
The calculation was driven by the total $gg$ cross section taking 
$\sigma_{gg} = (9/4) \sigma_{gq} = (9/4)^2 \sigma_{qq} = 3$ and 10 mb. 

At parton freezeout, GFR was applied using,
as common in transport approaches\cite{Nagle},
box Wigner functions
$W=\prod_{i,j} \Theta(x_m-|\vx_i-\vx_j|)\Theta(p_m-|\vp_i-\vp_j|)$,
with $x_m = 1$ fm.
This way (\ref{coaleq}) has a simple 
probabilistic interpretation: if $W=1$ 
(and the quantum numbers match) the hadron is formed, 
otherwise it is not ($W=0$).
When several coalescence final states were possible
for a given constituent, one was chosen randomly in an unbiased fashion.
Meson channels to $\pi$, $K$, $\eta$, $\eta'$,
$\rho$, $K^*$, $\omega$, $\Phi$;
and baryon channels to
$p$, $n$, $\Sigma$, $\Lambda$, $\Xi$,
$\Delta$, $\Omega$ were considered.
Gluons were split to a $q-\bar q$ pair, with asymmetric 
momentum fractions $x=0$ and $x=1$,
effectively corresponding to $1 g \to 1$ quark (or antiquark).
Easy color neutralization was also assumed.
Partons that did not find a coalescence partner were fragmented using
JETSET~7.4.10 \cite{JETSET}. JETSET was also used to decay unstable hadrons.

The parton initial conditions in \cite{v2} were utilized
but with LO pQCD minijet three-momentum distributions for $p_\perp > 2$ GeV ($K=2$, GRV98LO, $Q^2{=}p_T^2$),
smoothly extrapolated spectra below $p_\perp < 2$ GeV
to yield a total parton $dN(b{=}0)/dy=2000$ at midrapidity
(motivated by the observed $dN_{ch}/dy \sim 600$ and the expectation that
coalescence dominates the production),
and perfect $\eta=y$ correlation.

\section{Key results\protect\footnote{Due to page limitations, only a few key results are shown here. See \cite{slides} for the rest of the results.}}

\begin{figure}[hbpt] 
\begin{center}
\hspace*{0.1cm}\epsfig{file=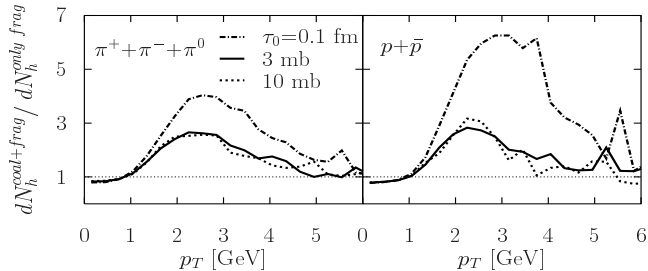,height=1.4in,width=3.35in,clip=5,angle=0}
\end{center}
\vspace*{-0.2cm} 
\caption{\label{fig1}
Pion and proton enhancement from parton coalescence
as a function of $p_\perp$ and parton cross section.
}
\end{figure}

Figure~\ref{fig1} shows the relative enhancement of pion and proton
production due to the coalescence process for $Au+Au$ at $b=8$ fm.
The ratio of the spectra from a calculation with
both coalescence and fragmentation hadronization channels
to that from a calculation with fragmentation only is plotted,
for $\sigma_{gg} = 3$ mb (solid curve),
10 mb (dotted), and a scenario with immediate freezeout (IFS) on the formation 
hypersurface (dashed-dotted).
In all three cases coalescence gives a large
enhancement for both species
in the intermediate $2 < p_\perp < 4{-}5$ GeV window,
in agreement with earlier expectations\cite{dukeCoal,texbudMtoB,coalv2}
based on (\ref{coaleq}).
However, unlike the earlier results,
for realistic $\sigma_{gg}=3-10$ mb,
the dynamical approach gives roughly the {\em same} enhancement for 
both $\pi$ and $p$, i.e., no additional enhancement for baryons. 
The reason for this is that, in the weak-binding case assumed,
baryons are more fragile: they have three constituents and therefore
less chance to escape without further interactions.
On the other hand, IFS, which ignores this dynamical effect,
enhances baryons over mesons.

\begin{figure}[hbpt] 
\begin{center}
\hspace*{-0.1cm}\epsfig{file=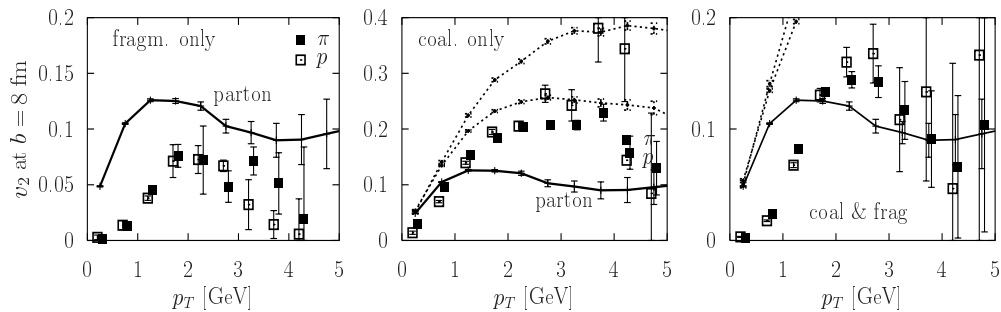,height=1.6in,width=5.2in,clip=5,angle=0}
\end{center}
\vspace*{-0.2cm} 
\caption{\label{fig2}
Coalescence and fragmentation contributions to elliptic flow (see text).
}
\end{figure}

Figure~\ref{fig2} shows pion and proton elliptic flow $v_2(p_\perp)$ 
for $\sigma_{gg}=10$ mb.
The {\em left panel} show that if {\em all} 
partons are hadronized via fragmentation,
the final hadron elliptic flow is less than that of partons (solid line)
because fragmentation quenches the spectra and also smears out the flow 
(jet width $\langle |j_\perp| \rangle > 0$).
The {\em middle panel} shows the influence of dynamical correlations on the flow
hadrons coming from coalescence. Pion(proton) $v_2$ is 
reduced by 20(40)\% relative to the simple flow scaling 
expectations\cite{Voloshincoal} (dotted lines). 
Therefore, parton $v_2(p_\perp)$ curves extracted from the hadron flows
would underpredict the real parton flow, and also differ from each other by $20$\%.
The {\em right panel} shows that when both hadronization 
channels are included, flow scaling is further violated because
the fragmentation contribution reduces elliptic flow.

The above findings demonstrate that coalescence is an
important hadronization channel.
However, the results also show that dynamical effects
are potentially large.
Further studies are needed to reveal
what it takes to preserve the basic 
features of the simple coalescence formulas.

\ack
Computer resources by the PDSF/LBNL 
and the hospitality of INT/Seattle where part of this work was done are gratefully acknowledged.
This work was supported by DOE grant DE-FG02-01ER41190.

\end{document}